\documentclass{aa}
\usepackage{epsfig}

\def\etal{{\rm et al.\thinspace}}
\def\eg{{\rm e.g.\ }}

\def\ie{{\rm i.e.\ }}
\def\cf{{\rm cf.\ }}

\def\spose#1{\hbox to 0pt{#1\hss}}
\def\ltsimm{\mathrel{\spose{\lower 3pt\hbox{$\sim$}}
	\raise 2.0pt\hbox{$<$}}}
\def\gtsimm{\mathrel{\spose{\lower 3pt\hbox{$\sim$}}
	\raise 2.0pt\hbox{$>$}}}
\def\Mdot{\hbox{${\dot M}$} \,}

\def\km{{\rm\thinspace km}}
\def\s{{\rm\thinspace s}}
\def\yr{{\rm\thinspace yr}}
\def\kmps{\hbox{${\rm\km\s^{-1}\,}$}}
\def\erg{{\rm\thinspace erg}}
\def\ergps{\hbox{${\rm\erg\s^{-1}\,}$}}
\def\Msol{\hbox{${\rm\thinspace M_{\odot}}$}}
\def\Msolpyr{\hbox{${\rm\Msol\yr^{-1}\,}$}}

\begin{document}
   
\title{High-Resolution X-ray Imaging of the Colliding Wind 
Shock in WR~147}

\author{J.M. Pittard\inst{1} \and I.R. Stevens\inst{2} \and 
        P.M. Williams\inst{3} \and A.M.T. Pollock\inst{4} \and
        S.L. Skinner\inst{5} \and M.F. Corcoran\inst{6,7} \and
        A.F.J. Moffat\inst{8,9}}

\institute{Department of Physics \& Astronomy, The University of Leeds, 
        Woodhouse Lane, Leeds, LS2 9JT, UK
 \and Department of Physics \& Astronomy, The University of Birmingham,
        Edgbaston, Birmingham, B15 2TT, UK
 \and Institute for Astronomy, University of Edinburgh, Royal Observatory,  
        Blackford Hill, Edinburgh EH9 3HJ, UK
 \and Computer \& Scientific Co. Ltd., 230 Graham Rd., Sheffield, S10 3GS, UK
 \and CASA, UCB 389, University of Colorado, Boulder, CO 80309, USA
 \and Universities Space Research Association, 7501 Forbes Blvd, Ste 206,
        Seabrook, MD 20706, USA
 \and Laboratory for High Energy Astrophysics, Goddard Space Flight Center,
        Greenbelt, MD 20771, USA
 \and D\'{e}partment de physique, Universit\'{e} de Montr\'{e}al, C.P. 6128, 
        Succ. Centre-Ville, Montr\'{e}al, QC, H3C 3J7, Canada
 \and Observatoire du mont M\'{e}gantic}
\offprints{J. M. Pittard, jmp@ast.leeds.ac.uk}

\date{Accepted April 8, 2002}

\abstract{
We analyze new high-resolution {\it Chandra} X-ray images of the 
Wolf-Rayet binary system WR147. This system contains a WN8 star
with an early-type companion located 0.6'' to its north, and is the
only known early-type binary with a separation on the sky large enough 
for the wind-wind collision between the stars to currently be resolved 
at X-ray energies. The 5~ksec {\it Chandra} HRC-I image provides the 
first direct evidence for spatially extended X-ray emission in an early-type 
binary system. The X-ray emission peaks close to the position of the radio bow 
shock and north of the WN8 star. A deeper X-ray image is
needed to accurately determine the degree of spatial extension, to 
exactly align the X-ray and optical/radio frames, and to determine whether
part of the detected X-ray emission arises in the individual stellar winds.
Simulated X-ray images of the wind-wind collision have a FWHM consistent
with the data, and maximum likelihood fits suggest that a deeper 
observation may also constrain the inclination and wind momentum ratio of
this system. However, as the WR wind dominates the colliding wind X-ray 
emission it appears unlikely that $\Mdot_{\rm OB}$ and $v_{\infty_{\rm OB}}$
can be separately determined from X-ray observations. We also note an 
inconsistency between numerical
and analytical estimates of the X-ray luminosity ratio of the stronger 
and weaker wind components, and conclude that the analytical results 
are in error.
}

\titlerunning{X-ray Imaging of the Colliding Wind Shock in WR~147}
\authorrunning{Pittard \etal}

\maketitle

\label{firstpage}

\section{Introduction}
\label{sec:intro}
Massive, early-type binaries with powerful stellar winds can generate
a complex region of shock-heated plasma with temperatures in excess 
of $10^{7}$~K when these winds collide (Prilutskii \& Usov \cite{PU1976};
Cherepashchuk \cite{C1976}). X-ray observations 
provide a direct probe of the conditions within the collision zone and 
also of the unshocked attenuating material along the line of sight 
in the system (\eg Stevens \etal \cite{SBP1992}; 
Pollock \etal \cite{Po1999}; Pittard \& Stevens \cite{PS1997}). 
Single early-type and Wolf-Rayet (WR) stars are also known 
to be X-ray emitters. The most viable model currently proposed is that the 
$10^{6} - 10^{7}$~K X-rays are generated by shocks in the wind resulting 
from the unstable nature of the line-driven acceleration 
(\eg Owocki \etal \cite{OCR1988}; Feldmeier \etal \cite{FPP1997}).
Direct spectroscopic evidence of clumps/shocks in WR winds was
first noted by Moffat \etal (\cite{MDLR1988}). 

Emission from colliding winds is often observable as a strong 
high-temperature X-ray excess
above that expected from the individual stars, together with phase-locked 
orbital variability resulting from the changing 
line of sight into the system (\eg Pollock \cite{Po1987}; 
Chlebowski \cite{C1989}; Williams \etal \cite{W1990}; 
Chlebowski \& Garmany \cite{CG1991}; 
Corcoran \cite{C1996}). Recent {\it Chandra} observations have revealed 
spectral features which are noticeably different to those from
single early-type stars, including strong forbidden line emission 
(Corcoran \etal \cite{CISP2001}; Pollock \etal \cite{Po2002}; 
Skinner \etal \cite{SGSS2001}).
In all previous X-ray observations, the colliding wind 
emission has remained spatially unresolved, and contaminated by 
the emission from the individual winds.

Non-thermal radio emission from WR stars is also a good indicator of wind-wind 
interaction (\eg Dougherty \& Williams \cite{DW2000}). Such emission is 
thought to be synchrotron radiation arising from electrons accelerated 
to relativistic velocities in the wind-wind collision zone
(Eichler \& Usov \cite{EU1993}). 
The presence of nonthermal radio emission from WR~147 (AS~431) was 
reported by Caillault \etal (\cite{C1985}) and Abbott \etal (\cite{ABCT1986}).
Later observations with {\it MERLIN} at 5~GHz resolved two radio sources
(WR~147S and WR~147N) separated by $\sim 0.6$\arcsec (Moran \etal 
\cite{M1989}). 
The southern thermal source (WR~147S) was coincident with the WR star, 
and the northern source (WR~147N) was proposed to be responsible for 
the observed nonthermal emission. Independent observations with the 
{\it VLA} confirmed this hypothesis (Churchwell
\etal \cite{C1992}; Skinner \etal \cite{SINZ1999}). 
Moran \etal (\cite{M1989}) further suggested that
WR~147 could have an unseen companion associated with the northern radio
source. 

WR~147 attracted the attention of many astronomers when radio, infrared and 
optical observations at high spatial resolution provided unambiguous 
evidence of its binarity (Williams \etal \cite{W1997}, hereafter W97; 
Niemela \etal \cite{N1998}, hereafter N98). New, higher resolution 
{\it MERLIN} observations by W97 revealed that
at 5~GHz both sources are elongated with sizes of $\sim 170 \times 253$~mas 
(WR~147S) and $\sim 267 \times 79$~mas (WR~147N), and that they had a 
separation of $575 \pm 15$~mas. The {\it UKIRT} infrared and {\it HST} optical 
images revealed for the first time a faint companion star located to the 
north of the nonthermal source associated with WR~147N. Crucially, the 
relative positions of
the sources revealed that the companion star, although close to the 
nonthermal radio emission, was located slightly ($\approx$ 60 mas) 
more distant from the WR star. The resulting picture is thus consistent
with the nonthermal emission arising in a colliding winds shock between the
two stars, with the wind of the WN8 star dominating.

The spectral type of the companion is currently somewhat uncertain, as the
excessive reddening towards this object ($A_{V} \approx 12$~mag; Churchwell
\etal \cite{C1992}) makes it hard to get a (blue) 
classification spectrum. Based on ram pressure arguments, a companion 
with a late-O or early-B spectral type is expected to have a sufficiently 
strong stellar wind to produce the bow-shock at its observed 
position\footnote{The latest attempt (Lepine \etal \cite{L2001}) 
based on STIS/{\it HST} resolved spectra yields an O5-7I(f) type, which seems a
little luminous for the observed magnitude difference.}. 
Although the mass-loss rate and wind velocity are not known for the OB star,
the wind momentum ratio can be directly determined from the geometry, 
specifically by comparison of the position of the nonthermal source with
those of the stars (\ie the ratio $r_{\rm OB}/D$, \cf Usov \cite{U1992}).
Estimates for the wind momentum ratio,

\begin{equation}
\label{eq:eta}
\eta = \frac{(\Mdot v_{\infty})_{\rm OB}}{(\Mdot v_{\infty})_{\rm WN}},
\end{equation}

\noindent range from $0.011^{+0.016}_{-0.009}$ (W97)
to $0.028^{+0.172}_{-0.027}$ (N98). This value can then
be used to estimate the wind momentum of the OB star if we know that of the
WN star (see Section~\ref{sec:wind_param}). The inclination angle can also
be estimated from the observed geometry: W97 deduced $i=41^{\circ}$, and
Contreras \& Rodr\'{i}guez (\cite{CR1999}, hereafter CR99) determined 
$i = 45 \pm 15^{\circ}$ from {\it VLA} observations. 

With WR~147 revealed as a colliding winds source it was expected that
most of the X-ray emission could be from the wind-wind collision.
A recent observation with {\it ASCA} (Skinner \etal
\cite{SINZ1999}) determined the emission to be thermal (although
multi-temperature) with the dominant component at $kT \approx 1$~keV. 
This is slightly harder than the characteristic temperature from single
early-type stars ($\approx 0.5$~keV). A high absorption column 
($N_{\rm H} \gtsimm 10^{22} \; {\rm cm^{-2}}$) in agreement 
with estimates based on the visual extinction was also confirmed. 
The intrinsic luminosity in the $0.5-10$~keV band was estimated as 
$L_{\rm x} = 3.5 \times 10^{32} \; \ergps$, which gives 
${\rm log\;L_{\rm x}/L_{\rm bol} \approx -6.5}$, 0.5~dex greater than the 
canonical value of $-7$ from single early-type stars.

WR~147 is the second closest WR star ($D \approx 650$~pc; Morris \etal 
\cite{M2000}) and one of a small number of WR systems where the 
companion stars have been spatially resolved (other examples being WR~86 and
WR~146 - N98). Of these, only WR~147 has a
sufficiently great separation between the stars for us to attempt to spatially 
resolve the wind-wind interaction region at X-ray energies. This in turn 
is currently only possible with the recently
launched {\it Chandra X-ray Observatory}, which combines a superb mirror with
sub-arcsecond resolution (van Speybroeck \etal \cite{vS1997}) and accurate, 
stable pointing. In this paper we report on the analysis of a 5~ksec 
observation of WR~147 with the High Resolution Camera (HRC-I). 

\begin{figure*}
\begin{center}
\psfig{figure=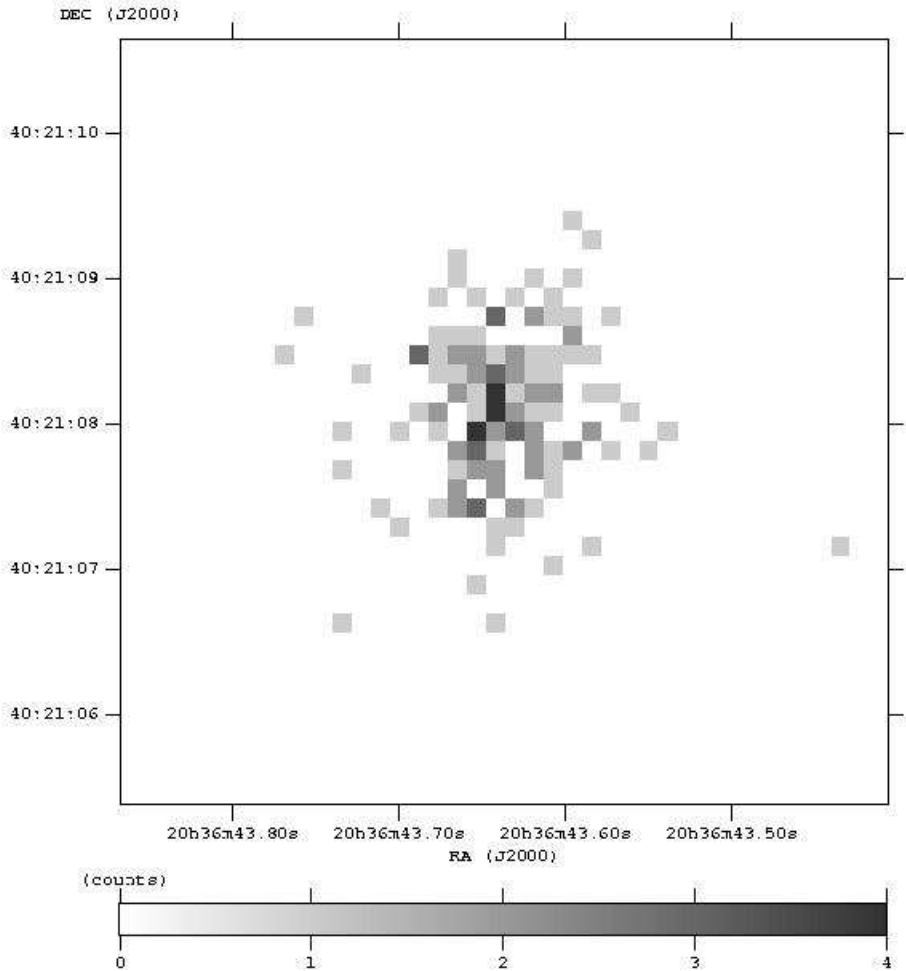,width=13.0cm}
\end{center}
\caption[]{Raw X-ray image of WR~147 containing 148 counts. There are
3 pixels towards the centre of the image which have the maximum of 4 
counts per pixel. Only 1 background count is expected in this image - all
of the others are from WR~147.}
\label{fig:xray}
\end{figure*}

\begin{table*}
\begin{center}
\caption{List of detected sources from the {\it wavdetect} algorithm. 
The extended emission at the position of WR~147 is detected with the 
greatest significance, and the quoted positions are from an unblocked image.
All uncertainties, including the count rates, are determined from the 
wavelet analysis.}
\label{tab:sources}
\begin{tabular}{llll}
\hline
\hline
Source & RA (J2000) & Dec (J2000) & Net counts \\
\hline
WR~147 & $20\;36\;43.637\pm{\rm 0.003s}$ & $+40\;21\;08.10\pm{0.04}$ & $148\pm12$ \\ 
Source~2 & $20\;36\;30.082\pm{\rm 0.010s}$ & $+40\;21\;23.05\pm{0.10}$ & $19\pm6$ \\
\hline
\end{tabular}
\end{center}
\end{table*}

\begin{figure*}[h]
\begin{center}
\psfig{figure=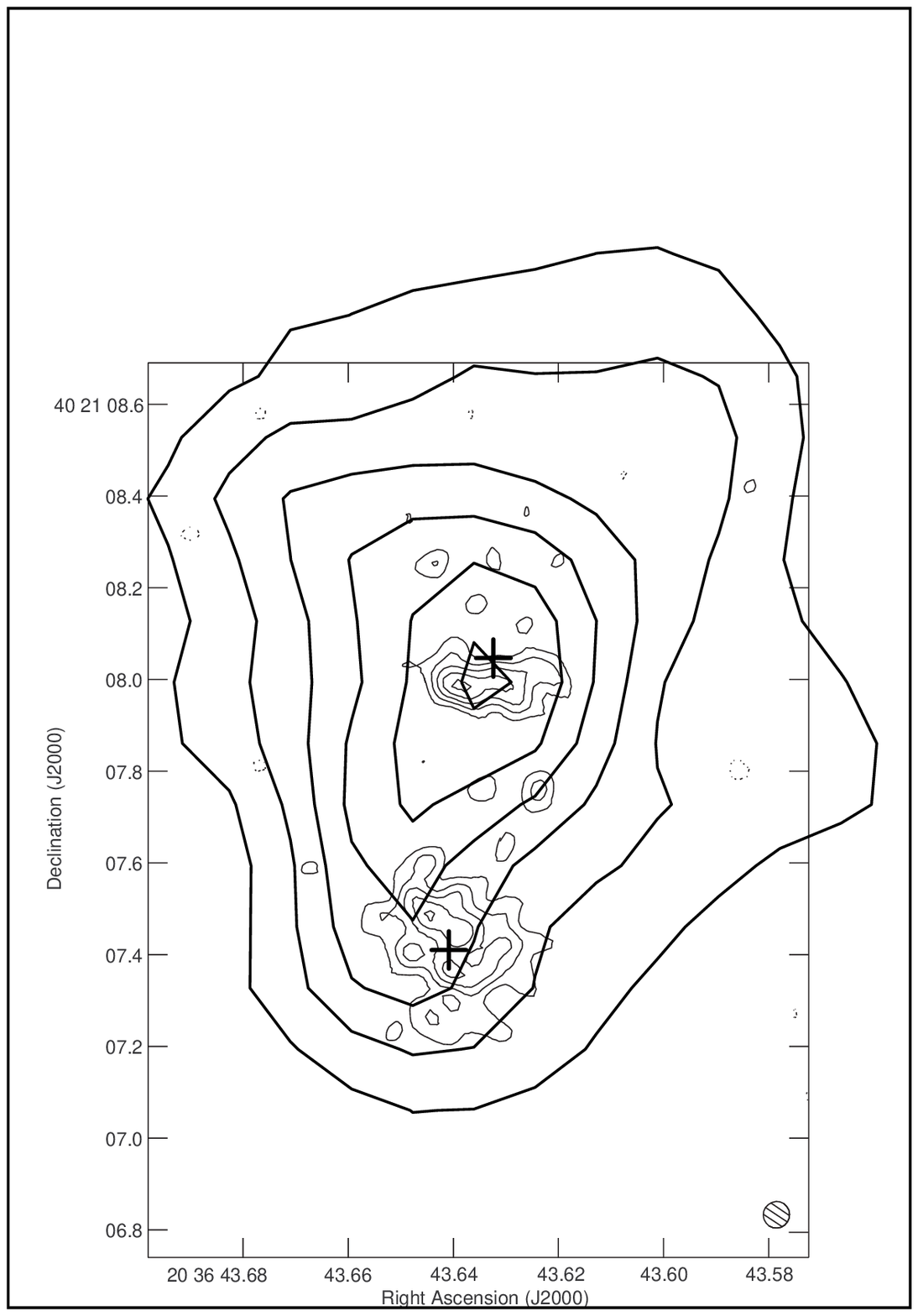,width=13.0cm}
%X-ray contours constructed from hrc_evt2_cntr_img2_csmooth1.fits (see line 97
%in commands file)
\end{center}
\caption[]{Contours (linearly spaced by 0.366 cts) from a smoothed 
HRC-I image of WR~147 (thick lines) 
overlaid onto high resolution radio contours from {\it MERLIN} (thin lines).
The positions of the WN8 (centroid of the thermal radio emission is
$20\;36\;43.64,\;\;+40\;21\;07.43$) and companion star (distance of 575~mas
and position angle of $351^{\circ}$ from the WR star) are marked with crosses.
Because the X-ray image could not be accurately
aligned with the optical/radio frame, there is some uncertainty involved 
in its absolute positioning. Despite this, the X-ray emission peak 
is certainly not cospatial with the WN8 star, and
seems likely to be cospatial with the nonthermal radio emission from the 
wind collision zone (although it is also consistent with the position 
of the companion star). Here we have applied an offset of 
(${\rm -0.00965s,\; -0.1387}$\arcsec) to the actual X-ray data 
(see Sec.~\ref{sec:analysis}).
Statistical tests reveal the X-ray emission to be
extended, and there is the possibility of some contaminating emission from
either or both stars (see discussion in Sec.~\ref{sec:syn_images}).}
\label{fig:xrayradio}
\end{figure*}

\section{HRC-I Analysis}
\label{sec:analysis}
{\it Chandra} (Weisskopf \etal \cite{WOvS1996}) observed WR~147 with 
the HRC-I instrument (Murray \etal \cite{M1997}) on 
{\mbox 2000 July 10 12:11 UT} 
to {\mbox 14:09 UT} for a total exposure time of 4.87~ksec. The HRC-I is a 
microchannel plate imager with excellent spatial and temporal
resolution but poor energy resolution. Each photon detected
by the HRC-I is time and position tagged. The size of the electronic readout 
is $6.429 \; \mu{\rm m}$ ($0.13175$\arcsec, referred 
to as `one pixel') which oversamples the point-spread-function (PSF). 
All {\it Chandra} data which have undergone
reprocessing have benefited from improvements made to the crossed-grid
charge detector signals. This has reduced the 50 per cent encircled 
energy radius from 4.0 to 3.2 pixels (0.527\arcsec~to 0.422\arcsec; 
Juda \etal \cite{J2000}). Softer photons are brought to a slightly
tighter focus than harder energies. Our analysis
is based on the pipeline processed data from the Rev1 Chandra X-ray Center 
(CXC) reprocessing.

The CXC has measured the on-orbit performance of
the Pointing Control and Aspect Determination (PCAD) system on {\it Chandra} 
(see Sec.~5.4 and Table~5.1 of the Proposer's Observatory Guide, 
hereafter POG), and determined that the standard processing is capable of
placing a reconstructed X-ray image on the celestial sphere to an accuracy of
$0.76$\arcsec (RMS) radius. The image was dithered during the observation
and these effects were removed during post-processing. We also checked that 
pointing transients had damped out by the time that the actual observation
began, and that the pointing remained stable during the observation.

A total of 115420 events were noted after the standard level 2 processing,
which yields a count rate across the entire detector of 
$23.7 \;{\rm cts\;s^{-1}}$. This implies a background rate of slightly less
than ${\rm 10^{-5}\;cts\;s^{-1}\;arcsec^{-2}}$. We therefore expect 
{\em only} 1 background count within the source area of WR~147 shown in 
Fig.~\ref{fig:xray}. Running the source detection software yielded 
one reliable source in addition to WR~147, ``Source 2'' in
Table~\ref{tab:sources}. We did not find an optical counterpart to this 
source in the {\it Tycho-2} catalogue from the {\it Hipparcos} 
mission (H\^{o}g \etal \cite{H2000}), but a counterpart was found
on images from the Palomar Sky Surveys (POSS). Its position has been 
measured on red and blue plates of the  
(original) POSS-I survey with the PMM machine for the USNO-A2.0 catalogue
(Monet \etal \cite{M1999}) to be $20\;36\;30.100,\;\;+40\;21\;23.32$~(J2000).
The offsets in RA and Dec from the {\it Chandra} detection are +0s.018 and 
+0.22\arcsec, close to the estimated astrometric precision (0.3\arcsec) of the 
USNO-A2.0 catalogue. The I magnitude of the coincident source is 
$\approx 12$, which is slightly too faint for {\it Tycho}.

We also measured the star's coordinates from 
{\it SuperCosmos} (\eg Hambly \etal \cite{H1998}) scans of glass copies 
of the POSS-II survey. It appears on the overlap region of 
two red plates giving coordinates $20\;36\;30.058,\;\;+40\;21\;22.97$
and $20\;36\;30.074,\;\;+40\;21\;22.91$ (J2000). These 
positions are independent of each other, coming from reductions using 
different sets of {\it Tycho2} stars centred about 5 degrees apart. The mean 
offsets from these coordinates to the {\it Chandra} Source 2 are -0.014s and
-0.16\arcsec, in the opposite sense to the POSS-I offsets. The sets of 
stellar coordinates from the POSS-I and POSS-II plates are consistent 
within the uncertainties of their
measurement but the difference (0.54\arcsec) might also reflect proper
motion of the star between the epochs of the POSS-I and POSS-II survey 
observations.

Finally a search in the 2MASS catalogue also 
revealed a possible match for the second source, although the positional
accuracy is too poor to help to register the HRC-I image. 
In any event, because we have only one X-ray source
additional to WR 147 in the field, which could be a line-of-sight 
coincidence with the star measured off the POSS, we will not use these
offsets to refine the alignment of the X-ray and optical/radio positions,
and must accept some astrometric uncertainty in our analysis of the 
{\it Chandra} data.

The accuracy of the radio positions of WR~147 and their tie to the 
optical frame comes via the phase reference source, and the uncertainty is not 
much more than 10~mas in each coordinate. On the other hand, 
the uncertainty in the X-ray positions relative to the optical frame is a 
combination of the source centroid on the detector (as listed in 
Table~\ref{tab:sources}), and of the tie of the X-ray to the optical frame. 
It is this last factor which dominates the uncertainty 
in the relative radio/X-ray positioning. As already noted, the standard 
processing is accurate in a statistical sense to 0.76\arcsec (RMS). 
However, the {\it Chandra} and {\it SuperCosmos}-POSS-II positions of 
source 2 suggests that the positioning of the X-ray frame is accurate 
to -0.014s and -0.16\arcsec, which is consistent with the uncertainties
in the position of the source centroids on the X-ray detector and also 
comparable to the {\it SuperCosmos} uncertainties. If we were to incorporate
this shift, the X-ray peak would lie on the western edge of the radio
bow shock. Hence the intrinsic astrometry appears to be much more
accurate for our observation than indicated by the RMS uncertainty, and we can
confidently state that the X-ray peak is definitely not cospatial with
the WR star. 

However, we would need to be able to align the X-ray and optical frames 
to a positional accuracy of better than 0.05\arcsec ($< 0.4$~pixels) 
to distinguish between the position of the radio bow shock or the companion 
star. This is clearly beyond the capability of our current dataset.
Therefore in Fig.~\ref{fig:xrayradio} we have added an offset
(${\rm -0.00965s,\; -0.1387\arcsec}$) to the actual X-ray data so that the 
peak emission matches up with the centre of the nonthermal radio emission 
from the bowshock. This offset is well within the nominal accuracy of the 
{\it Chandra} aspect solution. In summary the X-ray peak is certainly not
cospatial with the WR star, but could be cospatial with either the
nonthermal radio peak or the northern
optical component.

Another crucial question concerns whether the detected emission is
consistent with an extended object. To address this point we constructed 
several independent tests. We first extracted a PSF at the position of 
WR~147 using monochromatic source photons 
of energy 1.5~keV. This is a sensible approximation to the actual spectrum
which is sharply peaked at $\approx 1.5$~keV (Skinner \etal \cite{SINZ1999}).
We then extracted radially averaged surface brightness plots of the PSF 
and WR~147, and found that the full width half maximum (FWHM) of the former 
was $\approx 0.45\arcsec$ 
(in good agreement with the $0.4\arcsec$ quoted in the POG), compared to 
$\approx 1.0\arcsec$ for the WR~147 data.

In a second test we fitted a Gaussian function to the PSF and the HRC-I 
data of WR~147, and used a maximum likelihood method to determine the 
optimal FWHM. 
Despite experiencing complications in our fit to the PSF data due to its
extended wings, we again found that the FWHM of the WR~147 data 
($\approx 0.85\arcsec$) was significantly larger than that of the PSF 
($\approx 0.5\arcsec$). Finally, in a third test we investigated how 
the likelihood varied if we assumed that the
source was a circular region of constant surface brightness and adjusted its
radius. When fitting to the PSF, we found that the likelihood declined rapidly
as the radius was increased. In contrast, the likelihood was almost constant
for a source radius between 0 and 5 pixels when fitting to the WR~147 data.
(\ie for a circular source up to a radius of 0.66\arcsec). This contrast in 
behaviour again supports the inference that WR~147 is extended.
We therefore conclude that WR~147 {\em is} resolved at X-ray energies, 
and note that to the best of our knowledge this is the {\em only} stellar 
system where this is the case.

While the HRC-I has poor spectral resolution, previous {\it ASCA} data 
(Skinner \etal \cite{SINZ1999}) determined that there were essentially no 
X-ray photons from WR~147 with energies between 0.4-1.0~keV. Since the
extinction by dust towards WR~147 is very high, we conclude that
there essentially all X-ray photons with energies below 1.0~keV are absorbed.
It is therefore highly likely that (at least the majority of) the emission 
detected in the HRC-I is from the wind-wind collision zone, and we again 
note consistency with our conclusion that we are seeing extended 
emission from WR~147.

\begin{figure}
\begin{center}
\psfig{figure=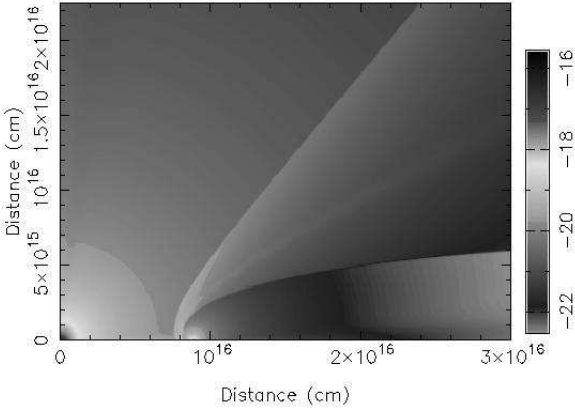,width=8.7cm}
\end{center}
\caption[]{Hydrodynamic simulation of the colliding winds in 
WR~147. The WN8 star is located at the bottom left of the plot at
position (0,0), with the companion at position ($8.8 \times 10^{15}$,0).
Shown is a gray-scale of the density 
(${\rm log_{10}} \rho \; {\rm g\;cm^{-3}}$).
It is clear that the collision region is globally stable. The parameters 
used in this model (cw\_2) are: 
$\Mdot_{\rm WN} = 1.5 \times 10^{-5} \; \Msolpyr$, 
$\Mdot_{\rm OB} = 2.8 \times 10^{-7} \; \Msolpyr$, 
$v_{\infty_{\rm WN}} = 950 \; \kmps$, $v_{\infty_{\rm OB}} = 1000 \; \kmps$.}
\label{fig:hydro}
\end{figure}

\section{Theoretical X-ray Emission Maps}
\label{sec:theory}
Based on the above conclusions, we now explore the feasibility of using
X-ray imaging to constrain some of the fundamental parameters of this 
system. This involves computing a grid of synthetic emission maps from 
hydrodynamical calculations of the wind-wind collision. Since there is some
small possibility that each star is also contributing to the observed
emission, we also attempt to determine the relative luminosities of each of
the three possible components (\ie each star plus the wind-wind collision).

\subsection{Estimates of the Wind Parameters}
\label{sec:wind_param}
To estimate the X-ray luminosity from each component, we first need
to obtain values for some essential parameters, such as the mass-loss
rates of the stars.

On account of its proximity, WR~147
has been extensively studied over different energy bands, and the
wind parameters of the WN8 star are well determined. Tight constraints
were derived by Morris \etal 
(\cite{M2000}) using ground-based optical and near-infrared observations, 
combined with high resolution space-based {\it ISO} observations. 
They obtained a clumping-corrected mass-loss rate of 
$\Mdot_{\rm WN} = 1.5 \times 10^{-5} \Msolpyr$ (with filling factor 
$f({\rm radio}) \sim 10\%$), substantially lower than 
derived from optical or radio observations under assumptions of homogeneity.
The terminal wind velocity, obtained from helium recombination 
lines and fine structure transitions in the infrared, was also revised 
downwards to $v_{\infty_{\rm WN}} \sim 950 \; \kmps$. 
The abundances of the WN8 wind (which affect our calculation of the
X-ray emission from the wind-wind collision in Section~\ref{sec:hydro_models})
are also fairly well determined (\cf Morris
\etal \cite{M2000}). From information contained in their paper we have
adopted the following relative abundance by mass (sum = 1): 
${\rm H}=0.09$, ${\rm He}=0.9$, 
${\rm C}=10^{-4}$, ${\rm N}=0.01$, ${\rm O}=8 \times 10^{-4}$, 
${\rm Ne}=1.2 \times 10^{-3}$, ${\rm Mg}=4.8 \times 10^{-5}$,
${\rm Si}=10^{-3}$, ${\rm S}= 2.1 \times 10^{-4}$, 
${\rm Ar}=2.4 \times 10^{-5}$, ${\rm Ca}=4.7 \times 10^{-5}$,
${\rm Fe}=2 \times 10^{-3}$, ${\rm Ni}=8.8 \times 10^{-6}$.
For the OB companion we assume solar abundances.

As for almost every colliding winds system, the wind 
parameters of the companion are not well determined. 
It is not even possible to place a reliable upper limit on 
the mass-loss rate of the wind of the companion star from the non-detection
of flux in the high resolution radio images, since the position of the star
is on the edge of the nonthermal source where the flux gradient is very steep.
Furthermore, the values of $\eta$ derived by W97 and N98
depend on the value of $r_{\rm OB}$ (\cf Usov \cite{U1992}), which is the
difference of two rather similar separations, each with observational errors,
and hence very uncertain. We therefore choose to average the results of 
W97 and N98 in the observational frame, and determine a projected stellar 
separation of 639~mas. Then using the projected radio separation deduced 
by {\it MERLIN} (575~mas), we obtain $\eta = 0.012^{+0.008}_{-0.005}$, 
independent of inclination. 

Given this wide range, we investigate the X-ray emission properties from
a series of colliding winds models with varying $\eta$ and inclination, 
$i$, as listed in Table~\ref{tab:models}. To reduce the number of free 
parameters we fix the terminal velocity of the companion star at 
$1000 \; \kmps$, although there is plenty of scope for a higher velocity. 
With all of the other parameters fixed, this leads to the mass-loss rate 
of the OB star varying directly with the chosen value of $\eta$. As it
is possible that the X-ray emission
from the wind-wind collision is somewhat contaminated by emission from the
individual winds, model X-ray luminosities lower than observed are to be 
preferred, and we have weighted our parameter space accordingly.

\subsection{X-ray Emission from the Stars}
\label{sec:xray_stars} 
We first attempt to estimate the X-ray luminosity 
from the WN and OB stars. There is a broad consensus that X-ray emission 
from early-type stars is generated from shocks within the stellar 
wind (\cf Owocki \etal \cite{OCR1988}; Feldmeier 
\etal \cite{FPP1997}). However, it is very difficult to make
reliable estimates of the intrinsic emission from each star 
as the variation in the observed luminosity 
between similar objects can be very large. Despite this, the following 
calculations give a rough impression of what we might expect for 
the intrinsic luminosity from each star. For OB stars, 
Sciortino \etal (\cite{S1990}) note a strong
correlation between the intrinsic X-ray emission and the wind momentum rate:

\begin{equation}
\label{eq:lx_ob}
{\rm log}\; L_{\rm x} = 0.53^{+0.10}_{-0.12} \; {\rm log} (\Mdot v_{\infty}) +
17.8^{+3.5}_{-2.7}
\end{equation}

\noindent Substitution of the values of $\Mdot$ and $v_{\infty}$ from 
Table~\ref{tab:models} gives $L_{\rm x} \approx 10^{32} \ergps$ 
(although with an uncertainty of $\pm$3 orders of magnitude). Taking the 
central values on the constants in Eq.~\ref{eq:lx_ob} 
(and assuming a 0.01-10.0 keV band) we find that this
is between 5 and 35\% of the intrinsic colliding winds emission from 
our models (see Section~\ref{sec:hydro_models}).

Estimates of the intrinsic X-ray luminosity from WN8 stars have been made
by Wessolowski (\cite{W1996}) and Pollock (\cite{Po1987}), and again the
variance between objects can be large. This is partially caused by 
large uncertainties due to the small exposure durations that many of these
stars were observed for. Therefore, perhaps the most accurate estimate
can be made from the {\it ROSAT} observations of WR~16 (HD~86161; 
WN8h, v=8.44) and WR~40 (HD~96548; WN8h, v=7.85), neither of which were
convincingly detected during WR~16's 7554s~PSPC exposure or WR~40's 45386s~HRI
exposure, giving $L_{\rm x} \ltsimm 10^{31} \ergps$ for WN8 stars.
Thus the OB companion is more likely than the WR star to contaminate 
the emission from the wind-wind collision zone.

\subsection{X-ray Emission from the Wind-Wind Collision}
\label{sec:hydro_models}
In this section we determine some of the basic properties of the wind 
collision region, using both analytical and numerical methods.
We use the results from previous observations as a starting point, and
therefore assume that the wind momentum ratio lies between $0.005-0.02$ 
and the orbital inclination between $30-60^{\circ}$ (\cf W97, N98, CR99). 
The full range of computed models is listed in Table~\ref{tab:models}.

First we use an analytical approximation (\cf Eichler \& Usov \cite{EU1993}) to
estimate the half-opening angle of the contact discontinuity, 
measured from the line between the secondary star and the shock apex.
For the models in Table~\ref{tab:models} we find that 
$\theta \simeq 31^{\circ}$ (models ${\rm cw\_1} - {\rm cw\_3}$),
$\theta \simeq 26^{\circ}$ (models ${\rm cw\_4} - {\rm cw\_6}$), and
$\theta \simeq 19^{\circ}$ (models ${\rm cw\_7} - {\rm cw\_9}$).

The combined kinetic power of the two winds for the parameters in 
Table~\ref{tab:models} is approximately constant at $4.3 
\times 10^{36} \; \ergps$ (as it is dominated by the WN8 wind for which
we assume fixed parameters). An indication of which wind dominates the X-ray
emission can be found (to first order) by evaluating the characteristic
cooling parameter, $\chi$, of each wind. To do this we simply make use of 
Eq.~8 in Stevens \etal (\cite{SBP1992}), while noting that WN8 abundances 
lead to a similar overall emissivity as solar abundances. Values for 
$\chi_{\rm WN}$ and $\chi_{\rm OB}$ are listed in Table~\ref{tab:models}, and 
since $\chi_{\rm WN}$ and $\chi_{\rm OB}$ are always $\gg 1$, the 
emission from the shocked winds is close to the adiabatic limit. In such
circumstances, the denser wind dominates the emission (\cf 
Myasnikov \& Zhekov \cite{MZ1993}), and we indeed find from 
hydrodynamical models that $L_{\rm cw,WN} \approx 20 L_{\rm cw,OB}$. 
In this regard, however, the results from hydrodynamical models are in 
conflict with estimates from the relevant equations in Usov (\cite{U1992}), 
which predict that $L_{\rm cw,WN} \sim L_{\rm cw,OB}$. This is a 
worrying revelation as these equations are widely used in the literature, 
and we investigate this discrepancy in a further paper (Pittard \& Stevens
\cite{PS2002}).

To obtain accurate estimates of the X-ray emission from the wind-wind
collision, and to generate synthetic emission maps with which to
compare to the {\it Chandra} data (see Section~\ref{sec:syn_images}),
we have computed hydrodynamical models of the wind collision,
calculated using {\sc VH-1} (Blondin \etal 
\cite{BKFT1990}), a Lagrangian-remap version of the third-order accurate
Piecewise Parabolic Method (PPM; Colella \& Woodward \cite{CW1984}).
The stellar winds were modelled as ideal gases with adiabatic index 
$\gamma = 5/3$. Although the wind collision is largely adiabatic, for
completeness we included radiative cooling via the method of operator
splitting. The cooling curve for the temperature range $4.0 < 
{\rm log}\; T < 9.0$ was generated using the Raymond-Smith plasma code
(Raymond \& Smith \cite{RS1977}), which specifies an effectively thin 
plasma in ionization equilibrium.

\begin{table*}
\begin{center}
\caption{Parameters for the computed colliding winds models. The wind 
parameters for the WN8 star are $\Mdot = 1.5 \times 10^{-5}\;\Msolpyr$ and
$v_{\infty} = 950\;\kmps$ (Morris \etal \cite{M2000}). The intrinsic
$0.5-10$~keV luminosity measured with {\it ASCA} is 
$3.5 \times 10^{32}\;\ergps$ (Skinner \etal \cite{SINZ1999}). Since there 
could be additional contributions from the individual stars,
colliding winds models with luminosities
slightly lower than the observed value are preferred. Our parameter space
study therefore incorporates this fact. Column 4 lists the inclination
of our line of sight into the system ($i = 0^{\circ}$ is pole on, 
$i = 90^{\circ}$ is in the orbital plane). Column 5 and 6 list the 
characteristic cooling parameter for the shocked WN8 and OB winds. Column 
7 and 8 list the intrinsic (\ie unabsorbed) X-ray luminosity from the 
colliding wind region in the $0.01-10.0$~keV and $0.5-10.0$~keV 
bands respectively, which is dominated by the WN wind. Column 9 
lists the number of counts in the synthetic images shown in 
Fig.~\ref{fig:comb_lincont}, assuming a 5~ksec exposure.}
\label{tab:models}
\begin{tabular}{ccccccccc}
\hline
\hline
Model & $\eta$ & $\Mdot_{\rm OB} \; (\Msolpyr)$ & $i$~(deg) & $\chi_{\rm WN}$ &
$\chi_{\rm OB}$ & $L_{(0.01-10.0)} \; (\ergps)$ & $L_{(0.5-10.0)} \; (\ergps)$ & HRC-I cts\\
\hline
cw\_1 & 0.02 & $2.8 \times 10^{-7}$ & 30 & 39 & 2560 & $2.0 \times 10^{33}$ & $ 4.6 \times 10^{32}$ & 99 \\
cw\_2 &      &                      & 45 & 48 & 3140 & $1.8 \times 10^{33}$ & $ 3.5 \times 10^{32}$ & 76 \\
cw\_3 &      &                      & 60 & 67 & 4430 & $1.2 \times 10^{33}$ & $ 2.6 \times 10^{32}$ & 55 \\
cw\_4 & 0.012 & $1.7 \times 10^{-7}$ & 30 & 39 & 4220 & $0.9 \times 10^{33}$ & $ 2.7 \times 10^{32}$ & 56 \\
cw\_5 &       &                      & 45 & 48 & 5170 & $ 1.0 \times 10^{33}$ & $ 2.2 \times 10^{32}$ & 45 \\
cw\_6 &       &                      & 60 & 67 & 7300 & $ 0.7 \times 10^{33}$ & $ 1.5 \times 10^{32}$ & 31 \\
cw\_7 & 0.005 & $7.1 \times 10^{-8}$ & 30 & 39 & 10100 & $ 0.6 \times 10^{33}$ & $ 1.1 \times 10^{32}$ & 20 \\
cw\_8 &       &                      & 45 & 48 & 12370 & $ 0.4 \times 10^{33}$ & $ 0.9 \times 10^{32}$ & 17 \\
cw\_9 &       &                      & 60 & 67 & 17470 & $ 0.3 \times 10^{33}$ & $ 0.6 \times 10^{32}$ & 12 \\
\hline
%\tablenotetext{\dag}{$(0.01-10.0\;{\rm keV})$}
%{\ddag}{$(0.5-10.0\;{\rm keV})$}
\end{tabular}
\end{center}
\end{table*}

The simulations were calculated with the assumption of cylindrical
symmetry - the orbital period of WR~147 is likely to be of order thousands
of years, so all orbital effects are negligible. 
The large orbital separation also implies that the 
winds are likely to collide at very near their terminal
velocities. We therefore assume that we can treat the winds as being
instantaneously accelerated to their terminal velocities at the stellar
surface of each star, and do not consider radiative driving effects.
For simplicity in this first investigation, we also assume that the
post-shock ion and electron temperatures equalize very rapidly. While the
equilibration timescale can be significant compared to the flow timescale
for wide systems (\eg WR~140 - see Zhekov \& Skinner \cite{ZS2000}), the 
exact situation is currently unclear with the importance of possible 
electron heating mechanisms in collisionless shocks still being debated.

Model cw\_5 lies in the middle of our grid of simulations with 
$\eta = 0.012$ and $i=45^{\circ}$. The latter specifies 
the orbital separation as $415/{\rm cos}\; i = 587 \; {\rm AU}$, and the
former the mass-loss rate of the companion as a reasonable 
$1.7 \times 10^{-7} \Msolpyr$. The computations were performed using grids 
spanning either ${\rm 400\;x\;300}$ cells (${\rm cw\_1} - {\rm cw\_6}$) or 
${\rm 600\;x\;400}$ cells 
(${\rm cw\_7} - {\rm cw\_8}$) with orbital separations of a minimum of 
125 or 290 cells respectively.

In Fig.~\ref{fig:hydro} we show the wind 
collision morphology for $\eta = 0.02$ and $i=45^{\circ}$. On account of 
its adiabaticity and approximately equal terminal velocities, the 
colliding winds region is stable to both thermal and Kelvin-Helmholtz
instabilities. The intrinsic emission calculated using Raymond-Smith 
emissivities (with appropriate abundances for the two winds) from model 
cw\_1 is within a factor of 2 of the measured emission
(see Table~\ref{tab:models}). Since the observed luminosity possibly
includes a significant contribution from intrinsic shocks in the wind of 
each star (\cf Owocki \etal \cite{OCR1988}), luminosities for the 
wind-wind collision which are slightly lower than the observed value 
are preferred.

\subsection{Synthetic Images}
\label{sec:syn_images}
To help to determine if the morphology seen in Fig.~\ref{fig:xrayradio}
is the result of colliding winds emission we have calculated theoretical X-ray 
images using the hydrodynamical models as input data. 

We would normally need to know which star is in front, as the stellar winds 
are themselves strong absorbers of X-rays. However, as a result of the long 
orbital period of WR~147, its orbital elements are unknown, and we do not
know which star is in front. Nevertheless, this is not problematical
because in WR~147 the absorption by the winds is actually very
low (a rough estimate of the wind absorption along a line of sight through 
the WR wind to the region of wind collision gives 
$N_{\rm H} \sim 2 \times 10^{19} \; {\rm cm^{-2}}$ \cf Usov \cite{U1992}). 
The low wind column is a direct consequence of the wide separation 
of the stars and the fact that the sightline to 
the wind collision region passes through the outermost, least dense regions 
of the winds. Therefore we can effectively ignore the exact direction of 
the line of sight into the system, which considerably simplifies the
calculation of synthetic images of the wind-wind collision.

As the absorption from spherical winds is negligible, we 
assume that the total column to WR~147 
($N_{\rm H} = 2.2 \times 10^{22} \; {\rm cm^{-2}}$; Skinner \etal 
\cite{SINZ1999}) is independent of orientation. It is well known that
the WR stars in Cyg~OB2 suffer more reddening than would be 
normally expected. A correlation between stellar luminosity and reddening
(Reddish \cite{R1967}) suggests that some of the reddening is circumstellar,
possibly left over from star formation. For the extinction we 
use Morrison \& McCammon (\cite{MM1983}) cross-sections.

As noted in the previous section, we find that the 
combined intrinsic spectrum is dominated by the shocked 
WN8 wind, so for simplicity we use WN8 emissivities for both winds. 
As $L_{\rm cw,WN} \gg L_{\rm cw,OB}$, this has
little effect on the resulting emission. We also incorporate 
the HRC-I effective area from data distributed with the PIMMS source code, 
and assume a distance 
of 650~pc (Morris \etal \cite{M2000}). The size in degrees of each cell
in the hydrodynamic grid varies with each model (\eg for model cw\_2
each cell is $2.142 \times 10^{-6}$ degrees square). Each HRC-I pixel is
$3.66 \times 10^{-5}$ degrees square, so for model cw\_2 each HRC-I pixel 
covers the same area of sky as 292 hydrodynamic cells. We have therefore
rebinned the synthetic images to the same scale as the HRC-I pixels, and have 
positioned the WR star at coordinates ($20\;36\;43.64,\;\;+40\;21\;07.4$), 
to match the centroid position of the radio contours in 
Fig.~\ref{fig:xrayradio}. The companion
star has been set at a position angle of $351^{\circ}$ from the WR-star
(\cf W97 and N98). All images are convolved with a Gaussian profile of 
$0.4\arcsec$ FWHM to approximate the HRC-I PSF. 

\begin{figure*}
\begin{center}
\psfig{figure=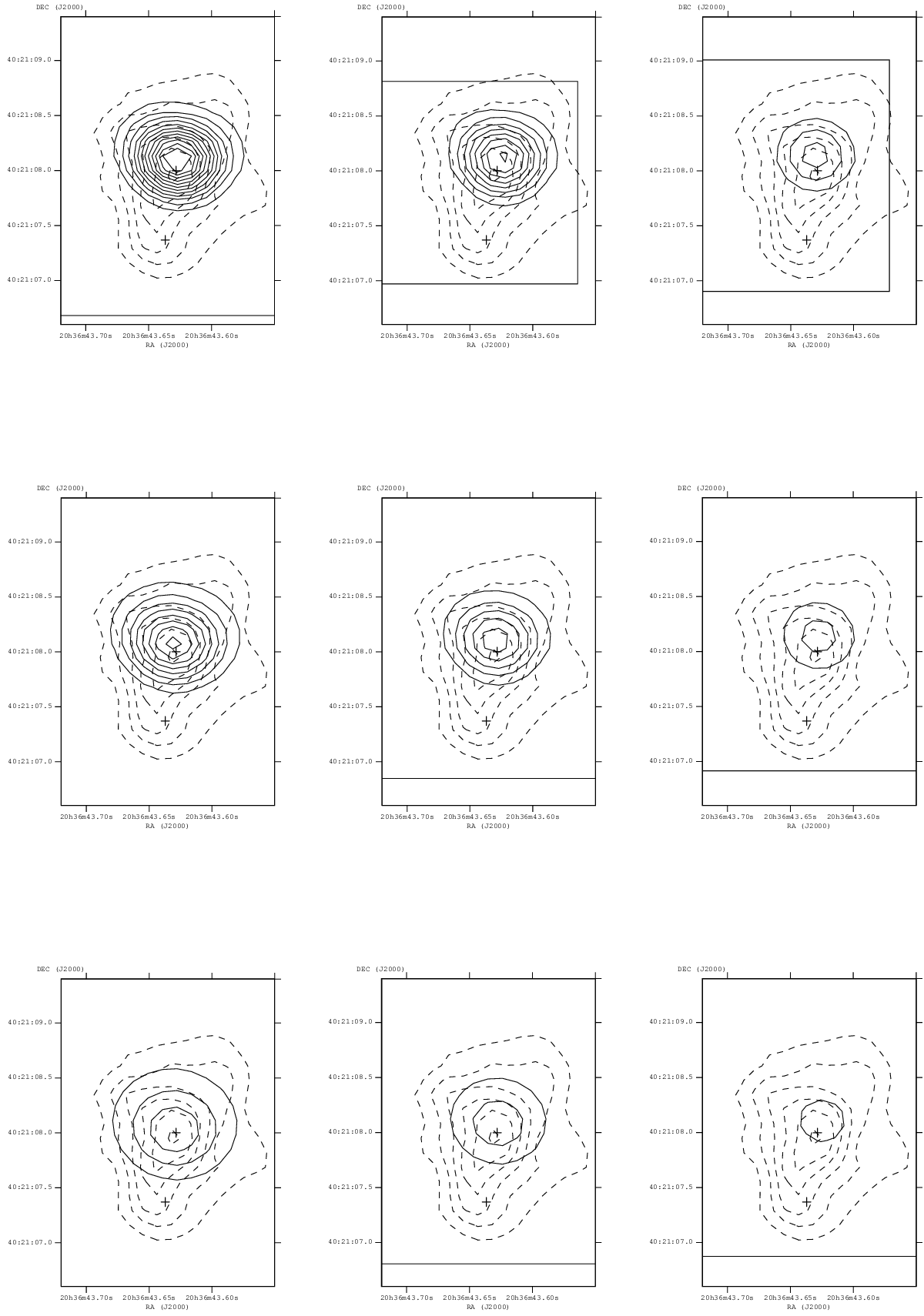,width=13.0cm}
\end{center}
\caption[]{Synthetic images (solid contours) of the colliding winds emission 
in WR147, generated from the hydrodynamic models listed in 
Table~\ref{tab:models}, and assuming a 5~ksec exposure (models are from top
to bottom, left to right, cw\_1 to cw\_9). The position of the
WR star has been set to match the centroid of the radio contours 
of the southern source in Fig.~\ref{fig:xrayradio}, and the companion star 
has been set at a position angle of $351^{\circ}$. The contours are linearly 
spaced by 0.366 cts to match the levels plotted in Fig.~\ref{fig:xrayradio}.
Images with more (less) than 6 contours have emission which is more (less) 
concentrated than the actual emission observed. The actual X-ray data 
(dashed contours) has been
shifted so that its peak lines up with the centre of the nonthermal
radio bowshock. The horizontal and 
vertical lines seen in some of the plots mark the edges of the computational
domain and should be discarded as artifacts. Values for ($i$, $\eta$) are 
given in the upper right corner of each image. $\eta$ decreases left to right,
and $i$ increases from top to bottom.}
\label{fig:comb_lincont}
\end{figure*}

Fig.~\ref{fig:comb_lincont} shows the synthetic images from each hydrodynamic
model assuming a 5~ksec exposure. For ease of comparison, each image has 
the same field of view as Fig.~\ref{fig:xrayradio} and the same contour 
levels (\ie contours linearly spaced by 0.366 cts). The number of counts
in each image is listed in Table~\ref{tab:models}. The contours 
of the actual observation in Fig.~\ref{fig:xrayradio} have a small additional 
broadening through our use of CSMOOTH, which is estimated to result in a total
net smoothing of approximately $0.42\arcsec$ FWHM. Therefore this should not 
overly affect the comparison. The offset mentioned in 
Sec.~\ref{sec:analysis} has again been applied.

\begin{figure*}
\begin{center}
\psfig{figure=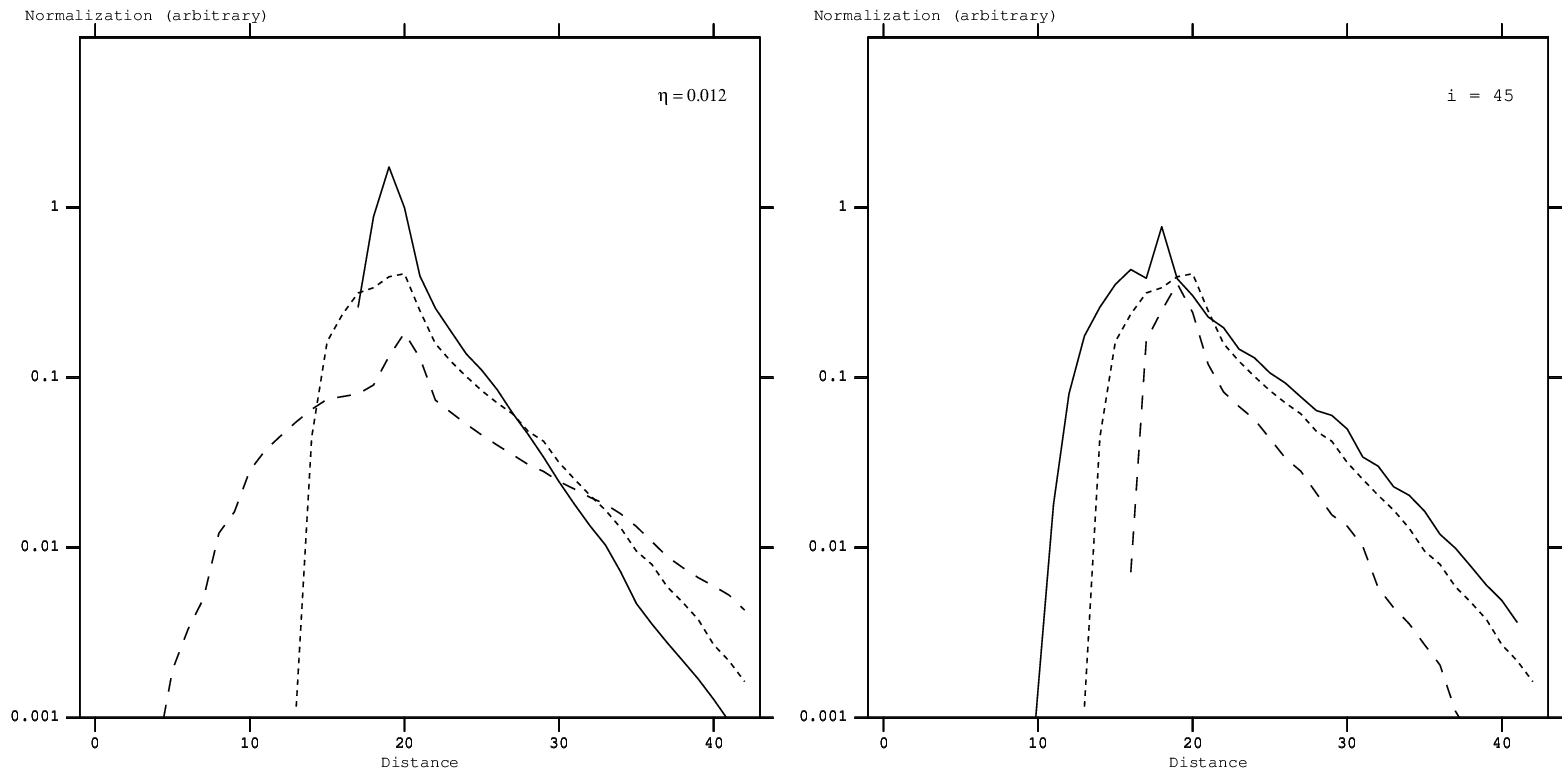,width=15.0cm}
\end{center}
\caption[]{Profiles through the shock cone along the line of centres of the
stars for selected synthetic images from Fig.~\ref{fig:comb_lincont}. 
Each unit on the x-axis corresponds to 31~mas (\ie 0.24 of a HRC-I pixel). The 
left panel displays the variation with $i$ ($i=30^{\circ}$ - solid; 
$i=45^{\circ}$ - dots; $i=60^{\circ}$ - dashes) for $\eta = 0.012$ 
(models cw\_4, cw\_5 and cw\_6 respectively). The right panel displays
the variation with $\eta$ ($\eta=0.02$ - solid; $\eta=0.012$ - dots; 
$\eta=0.005$ - dashes) for $i=45^{\circ}$ (models cw\_2, cw\_5 and cw\_8
respectively). A deep observation which could detect 
these differences might constrain values of $i$ and $\eta$.}
\label{fig:xeus_comb}
\end{figure*}

Of immediate note is the gratifying fact that the predicted FWHM 
from the synthetic images ($\approx 0.7-0.8\arcsec$) is in rough agreement 
with the value inferred from the data in Sec.~\ref{sec:analysis}. 
We also note that it appears to correlate with $\eta$ and $i$.
The range in model luminosity is apparent from the number and extent of 
the contour levels in each image, and it is evident that model cw\_1 
is too bright and model cw\_9 too faint. Because of the self-consistent 
generation of each model, the position of the peak emission lies at 
approximately the same coordinates in each image. However,
the contours appear more circular and more limb brightened for higher 
values of the system inclination, $i$ (see also Fig.~\ref{fig:xeus_comb}). 
For instance, models cw\_3 and cw\_4 have essentially the same count rate, 
yet the emission from model cw\_4 is much more concentrated as witnessed 
by its larger number of contours. 

The difference in the concentration of the emission between the 
models is best illustrated through profiles across the shock cone
as shown in Fig.~\ref{fig:xeus_comb}. The profile is clearly broader with
increased $\eta$ and increased $i$. While this width can only be used 
in comparison with observations to determine some combination of $\eta$ and 
$i$ (they cannot be determined independently from this comparison alone),
it is perhaps possible for these to be uniquely determined if further
additional comparisons between the models and the data are made. For instance,
the normalization of the profile is dependent on the X-ray luminosity from 
the colliding WN8 wind, since the shocked WR wind dominates the total 
colliding winds X-ray luminosity. With fixed values for $\Mdot_{\rm WN}$ 
and $v_{\infty_{\rm WN}}$, and since the shocked WN8 wind is largely 
adiabatic, $L_{\rm cw} \propto f/D$. Here $f$ is a measure of the shape of 
the wind-wind collision (and so is a function of $\eta$), and the separation 
of the stars, $D_{\rm sep}$, is related to the seperation on the sky, 
$D_{\rm sky}$, by $D_{\rm sep} = D_{\rm sky}/{\rm sin}\;i$. Hence
$L_{\rm cw}$ is a (complicated) function of $\eta$ and $i$. 

The width of the emission and the luminosity therefore
yield two constraints on two unknowns, and in principle $\eta$ and $i$ 
may be determined. We note that the observed circularity of the emission
also appears to vary, and it may be possible to use this as an additional
constraint in the determination of these parameters. 
Due to the weakness of the emission from 
the shocked OB wind, it is, however, clearly not possible to separately 
determine $\Mdot_{\rm OB}$ and $v_{\infty_{\rm OB}}$ from X-ray observations. 

From the contour maps in Fig.~\ref{fig:comb_lincont} we see that model cw\_5
has 5 contour levels, in good agreement with the number of levels in 
Fig.~\ref{fig:xrayradio}. However, the number of counts in the synthetic
image is only 45, approximately one third of the counts in the actual data.
Although the peak brightness is approximately correct, the synthetic image 
is significantly more compact than the actual data. In particular the 
synthetic emission does not extend as far south as the position of the 
WN8 star, whereas the actual X-ray emission clearly does. For comparison, the
lowest contour of model cw\_3 extends approximately the same distance in RA
as model cw\_5, but further in DEC (+40 21 07.58 to +40 21 08.59).

\begin{table}
\begin{center}
\caption{The assumed fractional emission from each of the three
components (wind-wind collision (WWC), WR wind, and OB wind) for the contour
maps in Fig.~\ref{fig:stars_comb}.}
\label{tab:var_em}
\begin{tabular}{cccc}
\hline
\hline
Panel & WWC & WR & OB \\
\hline
left   & 0.28 & 0.24 & 0.48 \\
middle & 0.28 & 0.36 & 0.36 \\
right  & 0.28 & 0.48 & 0.24 \\ 
\hline
\end{tabular}
\end{center}
\end{table}

The intrinsic luminosity of model cw\_5 is $\approx 63$\% of the {\it ASCA} 
value, whereas the number of simulated HRC-I counts for a 5~ksec exposure is
only $\approx 32$\% of the actual number of HRC-I counts. Therefore, 
the HRC-I count rate from models cw\_1 to cw\_9 is underestimated. This 
suggests that either: i) the simulated spectrum is too hard (from
comparing the collecting area of {\it ASCA} SIS with {\it Chandra} HRC-I
as a function of energy), or ii) simple fits to the {\it ASCA} SIS spectum 
return a value for the characteristic absorption column that is higher 
than the actual value. These two possibilities can each be influenced in
two ways. The spectrum could be softened by reducing the terminal
velocity of the WR star (since the emission from the shocked companion wind 
is insignificant its terminal velocity is irrelevant). However, this is
a fairly well defined parameter (see Morris \etal \cite{M2000}).
Alternatively, the spectrum could be softened if one or both stars 
contributed comparable luminosity (to the wind-wind collision) through
emission from intrinsic shocks in their winds, since it is well known that 
such emission is generally softer than emission from colliding winds.
Adressing point ii), we note that an overestimate of the absorbing 
column would be consistent with the degradation of the SIS detectors 
onboard {\it ASCA} (see Pittard \etal \cite{Pi2000}). One must
also be careful interpreting `characteristic values' when fitting single-
or two-temperature models to multi-temperature plasmas (\cf Strickland \& 
Stevens \cite{SS1998}). However, estimates from the visual extinction give 
$N_{\rm H} = 2.5\pm0.4 \times 10^{22} \;{\rm cm^{-2}}$, so there doesn't seem
to be much room to maneuver in this regard. 

It is tempting to conclude from the above discussion, and the fact that
the X-ray flux in Fig.~\ref{fig:xrayradio} is more 
extended in a north-south direction than predicted by any of the synthetic 
images in Fig.~\ref{fig:comb_lincont}, that the
most likely cause for this discrepency is the softening of the
X-ray spectrum by comparable intrinsic emission from the stars.
If true, and if the contamination is significant, it would seem that 
there is little hope that $\eta$ and $i$ can be determined.
While this would be disappointing for obvious reasons, we 
note that the S/N of the bottom data contour in Fig.~\ref{fig:xrayradio} is
very poor, and caution against this potential over-interpretation.
Since the absorption is so high towards WR~147, it is also expected that 
essentially all of the X-ray photons from the individual stars (which are
softer than those from the wind-wind collision) will be absorbed. On the 
other hand, it is indeed possible that we are detecting the few photons in 
the high energy tail of the stellar emission. 

We continue our analysis by performing a maximum likelihood fit 
of our models against the data. We find that the three models with 
$\eta = 0.02$ are most favoured, and that the general trend is for
$i = 60^{\circ}$ to also be favoured. Model cw\_3 appears
to be preferred over the others because its contours are
simultaneously broad (as suggested by the East-West extension of the 
data in Fig.~\ref{fig:xrayradio}) and not too concentrated 
(which, for example, counts against model cw\_1). The quality of
our data does not warrant further detailed fitting, but there is clearly
some promise that a future (deeper) observation could be used to
perform a consistency check on the best-fit parameter range determined by 
W97, N98, and CR99.

\begin{figure*}
\begin{center}
\psfig{figure=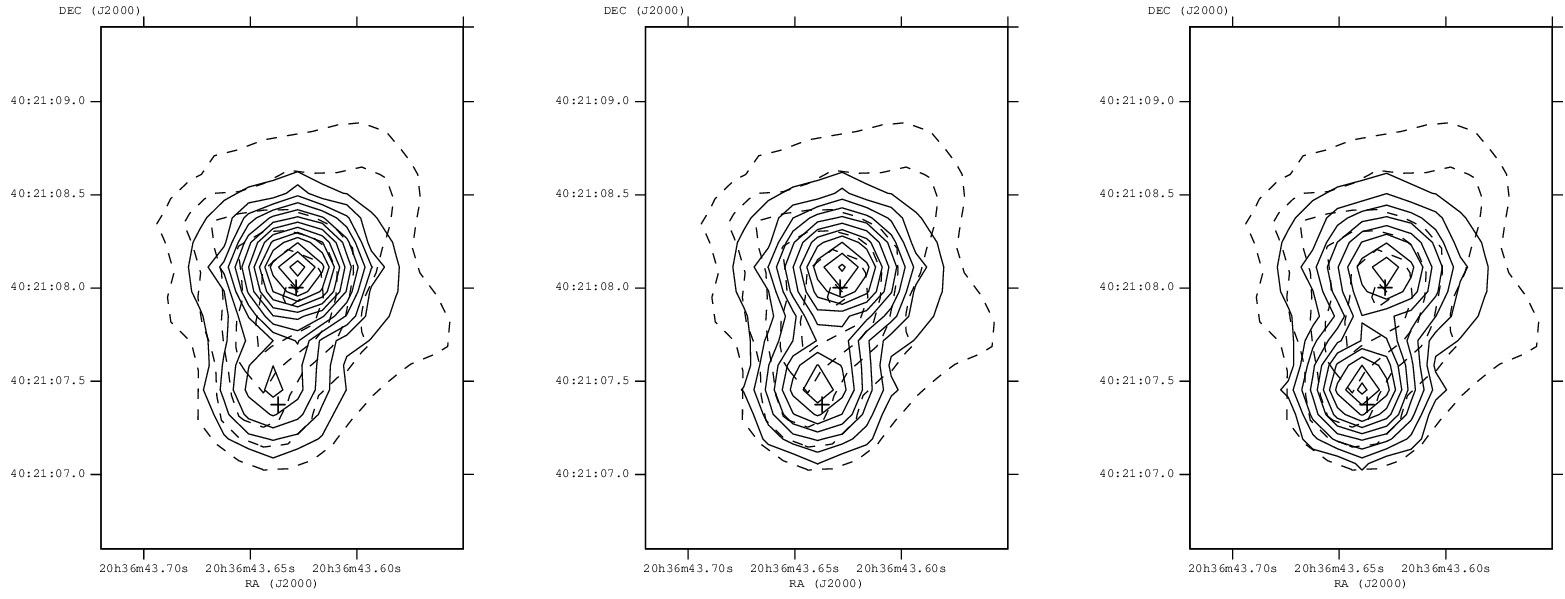,width=17.8cm}
\end{center}
\caption[]{Model cw\_3 with emission 
from the unshocked stellar winds included. In each panel the relative emission 
ratio between the three components is varied (see Table~\ref{tab:var_em}).
The contours are again linearly spaced by 0.366, and we have again 
superimposed the positions of the stars and shifted the actual X-ray 
contours.} 
\label{fig:stars_comb}
\end{figure*}

Since we are unable to completely rule out the possibility of contamination 
by stellar wind photons we have also examined the effect of adding X-ray 
emission at the positions of the stars, where the resultant 
simulated image has again been smoothed by a suitable Gaussian. 
Fig.~\ref{fig:stars_comb} shows the contours for differing ratios between 
the three components based on model cw\_3, with a small adjustment in the 
normalization of the predicted colliding winds emission. 
Table~\ref{tab:var_em} specifies the ratio of the individual contributions. 
Again the actual data and the position of the stars are superimposed.

It is immediately obvious from Fig.~\ref{fig:stars_comb} that the 
simulated emission is still too concentrated. However, it does become 
more extended, and
in particular the addition of emission from the WR star spreads the contours
much further south to give qualitative agreement with the actual data.
The addition of stellar contributions slightly improves the fitted 
likelihood in all 3 cases, although by a similar amount each time such that 
it is not clear which luminosity ratio is favoured.

The biggest problem between the data and our models is a lack of emission 
in our models at declinations greater than
approximately $+40\;21\;08.6$. This difference may be due to the 
incorrect alignment of the X-ray and radio data, the poor S/N of the lower
contour levels, the emission from the individual stellar winds extending
further from the stars than expected, or because our models do not
adequately represent the actual emission from the wind collision region.

\section{Conclusions}
\label{sec:conclusions}
The HRC-I X-ray maps presented in this paper show evidence for the 
{\em first} direct (\ie spatially resolved) detection of X-ray emission 
from the wind-wind collision zone of a massive early-type binary. The
data from WR~147 has a FWHM of $\approx 0.8\arcsec$, roughly double that
of the PSF. Models of the wind collision zone are also in rough quantitative 
agreement, and in particular predict a FWHM consistent with that measured.
To our knowledge this is the first time that X-ray emission has been resolved
in a stellar system.

A future exposure of longer duration is necessary to accurately align 
the X-ray and optical/radio frames, to potentially constrain the 
inclination and wind momentum ratio, to determine if the angular 
scale of the observed emission is consistent with models of the wind 
collision, and to determine the exact contribution of the individual stars
to the X-ray emission. 
Since the shocked WN8 wind dominates the colliding winds 
X-ray emission, it is unlikely that separate values for $\Mdot_{\rm OB}$ and 
$v_{\infty_{\rm OB}}$ can be determined from X-ray observations of this 
system.

Finally, we note the discovery that there is sometimes a discrepancy 
between estimates of the X-ray luminosity ratio of the shocked winds 
computed using the equations in Usov (\cite{U1992}) and calculations 
directly from hydrodynamic simulations.  While the ease of application 
of the elegant equations in Usov (\cite{U1992}) have led to their 
widespread use, this issue is examined in more detail in a further paper
(Pittard \& Stevens \cite{PS2002}).

%nd that therefore these 
%equations are unreliable. Values from the corresponding equations in the 
%radiative limit
%are also sometimes in error. Since the Usov equations are widely used in the 
%literature there is a clear need to quickly construct a suitable alternative.

\begin{acknowledgements}
JMP would like to thank PPARC for the funding of a PDRA position, and 
D. Strickland for use of his image generating program. We gratefully 
acknowledge the efforts of the {\it Chandra} team, in particular on the 
spectral resolution of the HRC and in determining the spatial extent of a 
detected source. We would
also like to thank the referee K. Gayley for his detailed critique, which
led to significant improvements in the final manuscript.
This work has made use of NASA's Astrophysics Data System Abstract Service.
\end{acknowledgements}

\end{document}